\documentclass[12pt,,letterpaper]{JHEP3}
\usepackage{graphicx}
\usepackage{epsfig}

\title{On thermal force from holographic action}

\author{Xiaomei Kuang\\
Center for Relativistic Astrophysics and High Energy Physics,\\
Department of Physics, Nanchang University, 330031 Nanchang,
China\\
\email{xmeikuang@gmail.com}}
\author{Yi Ling\\
Center for Relativistic Astrophysics and High Energy Physics,\\
Department of Physics, Nanchang University, 330031 Nanchang,
China\\
\email{yling@ncu.edu.cn}}
\author{Hongbao Zhang\\
Crete Center for Theoretical Physics, Department of Physics, \\
University of Crete, 71003 Heraklion, Greece\\
\email{hzhang@physics.uoc.gr} }
 \preprint{CCTP-2010-03}
 \abstract{Applying the relation between Euclidean on-shell action in the bulk and free energy on the
 holographic screen to a test charged particle in the charged R-N black hole, we show that not only
gravity but also electromagnetic force can be regarded as a sort of
thermal force, which accomplishes the unification of gravity and
electromagnetism to some extent. In addition, taking into account
the fact that both the temperature and thermal force of the dual
system are measured at infinity, we argue that the dual holographic
screen may be located at infinity.}
\begin{document}
\section{Introduction}
Inspired by the emergent phenomenon in the holographic scenario,
Erik Verlinde proposed that the origin of gravity can be interpreted
as a thermal force caused by changes in the information associated
with the positions of material bodies, i.e.,
\begin{equation}
f=T\frac{\Delta S}{\Delta x},\label{entropic}
 \end{equation}
 where $T$ and $S$ are the temperature and entropy related to the holographic screen, respectively\cite{Verlinde}.
 Since it is related to variation of entropy, this thermal force is usually called entropic
force. Such an entropic force proposal has stimulated much effort on
its further clarifications and possible
applications\cite{CCO1,Smolin,SG,Makea,CM,LW1,Gao,ZGZ,Wang1,Wang2,WLW,LW2,LKL,Zhao1,Myung,Jerzy,LWW,CCO2,Pesci,TW,MK,VS,Konoplya,Kar}.

In particular, recently Yue Zhao provides a clear-cut support for
such a hand-waving proposal by exploiting the relation between
on-shell Euclidean action of gravity theory and partition function
for the dual theory\cite{Zhao2}.

In this paper, employing such a holographic duality, we shall
further show that gravity and electromagnetic force experienced by a
charged particle in the charged R-N black holes can be unified as a
sort of thermal force\footnote{ Electromagnetic force as a thermal
force is also reached by Tower Wang\cite{Wang}, but the methodology
is different.}. As a bonus, our results also suggests that the dual
holographic screen may be located at infinity.

Before proceeding, it is noteworthy that the above thermal formula
(\ref{entropic}) is only valid for the
 micro-canonical ensemble. When the canonical ensemble is concerned,
 it follows from thermodynamics that the thermal force should be replaced by
 \begin{equation}
 f=-\frac{\Delta F}{\Delta x},\label{thermal}
 \end{equation}
 where $F$ is the corresponding free energy. In addition, Natural
 Units will be employed here, i.e., $k_B=G=c=\hbar=1$.
\section{Thermal force from holographic action}
Start from the charged R-N black hole
\begin{equation}
ds^2=g_{tt}dt^2+g_{rr}dr^2+r^2d\Omega^2,
\end{equation}
where
\begin{eqnarray}
g_{tt}&=&-(1-\frac{2M}{r}+\frac{Q^2}{r^2}),\nonumber\\
g_{rr}&=&=(1-\frac{2M}{r}+\frac{Q^2}{r^2})^{-1}
\end{eqnarray}
with $M$ and $Q$ the mass and charge of black hole, respectively.
The corresponding electromagnetic potential reads
\begin{equation}
A_a=-\frac{Q}{r}(dt)_a.
\end{equation}
Now consider a test particle
of mass $m$ and charge $q$ in such a background with the action
given by
\begin{equation}
S_p=\int
d\lambda(-m\sqrt{-g_{\mu\nu}\frac{dx^\mu}{d\lambda}\frac{dx^\nu}{d\lambda}}+qA_\sigma\frac{dx^\sigma}{d\lambda}),
\end{equation}
which does not depend on the choice of parameter $\lambda$. In what
follows, we will choose $\lambda=t$. In addition, in order to hold
the dual system in equilibrium, we also require the worldline of
this charged particle to be the orbit of the time-like Killing
vector field $(\frac{\partial}{\partial t})^a$. Thus the on-shell
action of the particle can be written as\footnote{Note that in later
calculations the zero component of four potential $A_t$ will be
regarded as a scalar field.}
\begin{equation}
S_p=\int dt(-m\sqrt{-g_{tt}}+qA_t).
\end{equation}

To proceed, we first go to the Euclidean section from the Lorentzian
section by setting $\tau=it$. Then the Euclidean R-N black hole is
given by
\begin{equation}
ds^2=g_{\tau\tau}d\tau^2+g_{rr}dr^2+r^2d\Omega^2
\end{equation}
with $g_{\tau\tau}=-g_{tt}$. As is well known, to regularize the
conical singularity appearing in this section, $\tau$ need to have a
period of $\beta$ given by
\begin{equation}
\beta=2\pi\frac{\sqrt{g_{rr}}}{\partial_r\sqrt{g_{\tau\tau}}}|_{horizon}=4\pi\frac{\sqrt{g_{\tau\tau}g_{rr}}}{\partial_rg_{\tau\tau}}|_{horizon},
\end{equation} which
actually corresponds to the inverse of temperature $T$ of the dual
system, measured at infinity. Since the back-reaction of the test
particle is neglected here, the temperature of the dual system keeps
unchanged, which implies the system should be described by the
canonical ensemble. Now by the holographic duality, the free energy
of the dual system is related to the on-shell Euclidean action of
the gravity system as
\begin{equation}
F=T(S_G+S_p),
\end{equation}
where the Euclidean action of the test particle $S_p$ is given by
\begin{equation}
S_p=\int
d\tau(-m\sqrt{-g_{tt}}+qA_t)=\frac{-m\sqrt{-g_{tt}}+qA_t}{T},\label{paction}
\end{equation}
and $S_G$ is the holographic action for the Euclidean background,
independent of the position of the test
particle\cite{KLS,Skenderis,MM,LS}. Thus Eq.(\ref{thermal}) follows
that the thermal force induced by the change of position of the test
particle can be written as
\begin{equation}
f^{th}_a=-\nabla_aF=\nabla_a(m\sqrt{-g_{tt}}-qA_t)=-(\frac{m}{2}\frac{\partial_rg_{tt}}{\sqrt{-g_{tt}}}+q\partial_rA_t)(dr)_a.
\end{equation}

On the other hand, taking into account that the total force and
electromagnetic force experienced by the test particle
read\cite{Wald}
\begin{eqnarray}
f^{tot}_a&=&m\frac{1}{\sqrt{-g_{tt}}}\nabla_a\sqrt{-g_{tt}},\nonumber\\
f^{em}_a&=&qF_{ab}\frac{1}{\sqrt{-g_{tt}}}(\frac{\partial}{\partial
t})^b=q\frac{1}{\sqrt{-g_{tt}}}\nabla_aA_t,
\end{eqnarray}
we can obtain the external force exerted on the test particle as
\begin{equation}
f^{ex}_a=f^{tot}_a-f^{em}_a=-f^{g}-f^{em}=\frac{1}{\sqrt{-g_{tt}}}f^{th}_a,
\end{equation}
where we have used $f^{g}=-f^{tot}$ when we view the curved
spacetime as an effect of force induced by gravity. Note that the
thermal force differs from the force exerted locally by the redshift
factor, so it can be identified as the force exerted at
infinity\cite{Wald}, i.e.,
\begin{equation}
F^{th}=F^{ex}_\infty,
\end{equation}
which along with the temperature of the dual system suggests that
the dual holographic screen where the dual system lives may be
located at infinity. In addition, if we further associate the
gravity and electromagnetism relevant free energy with the first and
seconde term in Euclidean action (\ref{paction}), then the above
equality implies that not only gravity but also electromagnetic
force can be regarded as a sort of thermal force. In this sense, the
unification of gravity and electromagnetism is realized on the dual
holographic screen, although it does not mean that we does not need
quantum gravity in the bulk, as we have had quantum electrodynamics.
\section{Discussions}
Applying the holographic duality, we have demonstrated that gravity
and electromagnetic force can be unified as a sort of thermal force
on the dual screen. In addition, both the temperature and thermal
force suggest that the dual holographic screen may be located at
infinity.

Although we focus ourselves onto the charged R-N black holes for
simplicity, our result should be applicable to general K-N black
holes. In addition, note that the temperature drops out in our
calculation of the thermal force, thus it is interesting to
investigate whether the thermal force interpretation may also be
generalized to arbitrary asymptotically flat stationary spacetime
with stationary electromagnetic field, even more general spacetime.
On the other hand, since the weak and strong force share the same
structure as the electromagnetic force, it is not difficult to
imagine that the weak and strong force can also be interpreted as a
sort of thermal force where the concept of force is suitable for
use. We expect to report all of these issues in the near future.
\section*{Acknowledgements}
HZ is grateful to Yue Zhao for helpful correspondence and Xin Gao
for stimulating discussions. He also thanks the wonderful
hospitality from Center for Relativistic Astrophysics and High
Energy Physics at Nanchang University, where this work is performed.
XK and YL are partly supported by NSFC(No.10875057), Fok Ying Tung
Education Foundation( No.111008), the key project of Chinese
Ministry of Education(No.208072) and Jiangxi young
scientists(JingGang Star) program. They are also supported by the
Program for Innovative Research Team of Nanchang University. HZ was
partially supported by a European Union grant
FP7-REGPOT-2008-1-CreteHEP Cosmo-228644 and a CNRS PICS grant \#
4172.


\begin{thebibliography}{10}
\bibitem{Verlinde}E. Verlinde, arXiv:1001.0785[hep-th].
\bibitem{CCO1}R. Cai, L. Cao, and N. Ohta, arXiv:1001.3470[hep-th].
\bibitem{Smolin}L. Smolin, arXiv:1001.3668[gr-qc].
\bibitem{SG}F. Shu and Y. Gong, arXiv:1001.3237[gr-qc].
\bibitem{Makea}J. Makea, arXiv:1001.3808[gr-qc].
\bibitem{CM}F. Caravelli and L. Modesto, arXiv:1001.4364[gr-qc].
\bibitem{LW1}M. Li and Y. Wang, arXiv:1001.4466[hep-th].
\bibitem{Gao}C. Gao, arXiv:1001.4585[hep-th].
\bibitem{ZGZ}Y. Zhang, Y. Gong, and Z. Zhu, arXiv:1001.4677[hep-th].
\bibitem{Wang1}Y. Wang, arXiv:1001.4786[hep-th].
\bibitem{Wang2}T. Wang, arXiv:1001.4965[hep-th].
\bibitem{WLW}S. Wei, Y. Liu, and Y. Wang, arXiv:1001.5238[hep-th].
\bibitem{LW2}Y. Ling and J. Wu, arXiv:1001.5324[hep-th].
\bibitem{LKL}J. Lee, H. Kim, and J. Lee, arXiv:1001.5445[hep-th].
\bibitem{Zhao1}L. Zhao, arXiv:1002.0488[hep-th].
\bibitem{Myung}Y. Myung, arXiv:1002.0871[hep-th].
\bibitem{Jerzy}J. Kowalski-Glikman, arXiv:1002.1035[hep-th].
\bibitem{LWW}Y. Liu, Y. Wang, and S. Wei, arXiv:1002.1062[hep-th].
\bibitem{CCO2}R. Cai, L. Cao, and N. Ohta, arXiv:1002.1136[hep-th].
\bibitem{Pesci}A. Pesci, arXiv:1002.1257[gr-qc].
\bibitem{TW}Y. Tian and X. Wu, arXiv:1002.1275[hep-th].
\bibitem{MK}Y. Myung and Y. Kim, arXiv:1002.2292[hep-th].
\bibitem{VS}I. V. Vancea and M. A. Santos, arXiv:1002.2454[hep-th].
\bibitem{Konoplya}R. A. Konoplya, arXiv:1002.2818[hep-th].
\bibitem{Kar}S. Kar {\it{et al}}., arXiv:1002.3976[hep-th].
\bibitem{Zhao2}Y. Zhao, arXiv:1002.4039[hep-th].
\bibitem{KLS}P. Kraus, F. Larsen, and R. Seibelink, Nucl.Phys.B563:259(1999).
\bibitem{Skenderis}K. Skenderis, Class.Quant.Grav.19:5849(2002).
\bibitem{MM}R. Mann and D. Marolf, Class.Quant.Grav.23:2927(2006).
\bibitem{LS}T. Liko and D. Sloan, Class.Quant.Grav.26:145004(2009).
\bibitem{Wald}R. Wald, Genereal Relativity(The University of Chicago
Press, Chicago, 1984).
\end{thebibliography}
\end{document}